\documentclass[twocolumn, a4paper]{revtex4}
\usepackage{graphicx}
\usepackage{dcolumn}
\usepackage{amsmath}
\usepackage{color}
\begin{document}
\hsize\textwidth\columnwidth\hsize\csname@twocolumnfalse\endcsname
\title{Single-Shot Readout with the Radio Frequency Single Electron Transistor in the
Presence of Charge Noise} \vspace{-0.5cm}

\author{T. M. Buehler$^{1,2}$, D. J. Reilly$^{1,2}$, R. P. Starrett$^{1,2}$,
Andrew D. Greentree$^{1,2}$,\\ A. R. Hamilton$^{1,2}$, A. S.
Dzurak$^{1,3}$ and R. G.Clark$^{1,2}$}

\affiliation{$^1$Centre for Quantum Computer Technology,
University of New South Wales, Sydney 2052, Australia}

\affiliation{$^2$School of Physics, University of New South Wales,
Sydney 2052, Australia}

\affiliation{$^3$School of Electrical Engineering \&
Telecommunications, University of New South Wales, Sydney 2052,
Australia}

\begin{abstract}

\noindent \small{ The radio frequency single electron transistor
(rf-SET) possesses key requirements necessary for
reading out a solid state quantum computer. This work explores the
use of the rf-SET as a single-shot readout device in the presence
of $1/f$ and telegraph charge noise. For a typical spectrum of
$1/f$ noise we find that high fidelity, single-shot measurements are possible for signals $\Delta q > 0.01e$. 
For the case of telegraph noise,
we present a cross-correlation measurement technique that uses two
rf-SETs to suppress the effect of random switching events on
readout. We demonstrate this technique by monitoring the charge
state of a metal double dot system on microsecond time-scales.
Such a scheme will be advantageous in achieving high readout
fidelity in a solid state quantum computer.}\end{abstract}
\maketitle

Electrical circuits that exploit entangled quantum states
have the potential to revolutionize information processing. One of
the key requirements for the realization of such technology is the
ability to perform reliable initial state preparation and readout
of quantum bits (qubits) with both high fidelity and efficiency
\cite{Nielsen_chuang}. This formidable challenge amounts to the
detection of single charge, flux \cite{Makhlin} or
spin\cite{Kane_nature,Vrijen} states on time-scales shorter than
the qubit mixing time using detectors that operate with
sensitivities near the quantum limit
\cite{caves,schoelkopf_nature}. The radio frequency single electron
transistor (rf-SET) is a readout device that possesses such requirements, exhibiting high
sensitivity, {\it switchable} back-action \cite{Johansson_PRL} and
a quantum measurement efficiency that approaches unity
\cite{clerk_PRL}.

As pioneered by Schoelkopf {\it et al.,} \cite{Schoelkopf_science},
the rf-SET consists of an SET electrometer \cite{natobook}
embedded in an impedance matching network to enable fast
operation. SETs are sensitive to charge since current flow through
the device depends critically on the electrostatic potential of
the center `island'. Here we discuss `single-shot' 
measurements in which the rf-SET is used to detect a charge state in a single pass, without 
repeating the preparation of that state. Although single-shot measurements are the most 
efficient mode of readout (and desirable for quantum error correction), they are challenging in the case of fast operation since bandwidth-narrowing
techniques commonly used to reduce noise also limit the response time. 
 
The purpose of the present work is to address how single-shot
measurements are affected by {\it charge noise}. In our setup we
distinguish between two types of charge noise: $1/f$ noise arising
from many fluctuating charge traps that are weakly coupled to the
SET and telegraph noise due to single traps that are strongly
coupled to the SET \cite{weissman}. We find that the effect of $1/f$ noise
on read-out fidelity is negligible in our set-up for measurement times
shorter than $\mathrm{t_{meas} \approx 10 \mu s}$. However telegraph noise, characterized by discontinuous switching of charge between
metastable trap configurations can affect readout fidelity since switching may occur on timescales commensurate with the 
measurement time.

To investigate the effects of
telegraph noise on readout fidelity we have built an
$\mathrm{Al/Al_{2}O_{3}}$ readout simulation device, where the
charge state of two coupled metal dots is measured with a twin
rf-SET detector. We present single-shot measurements of this nano-scale device
taken simultaneously with two rf-SETs. These results demonstrate the feasibility 
of using rf-SETs to perform readout in a single-shot, via the use of a cross-correlation 
technique to greatly suppress the effect of charge noise. Such measurement 
schemes maybe important if rf-SETs are to achieve high 
fidelity readout, in the presence of charge noise.

\begin{figure}
\begin{center}
\includegraphics[width=6cm]{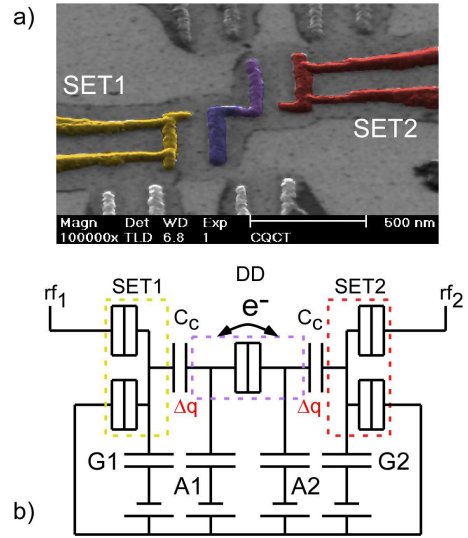}\vspace{-0.3cm}
\caption{ \small{ \textbf{a)} Electron micrograph of
the twin rf-SET and double-dot  device. The rf-SETs detect the charge state of the
two metal dots which are coupled by a tunnel junction (center of image).  \textbf{b)} Is a schematic showing the two rf-SETs and the double-dot with associated gates and 
capacitances.
}}\label{fig_intro} \vspace{-0.9cm}
\end{center}
\end{figure}

Turning now to the details of our experiment, 
Figure 1 shows an electron micrograph (a) and schematic (b) of our
device, including gates A1,A2,G1,G2 and the double-dot structure (single tunnel barrier) capacitively coupled to the two rf-SETs
either side. Simultaneous independent operation of two rf-SETs is made possible
using the technique of wavelength division multiplexing in
conjunction with a single cryogenic amplifier. Details of our
setup can be found in Ref. \cite{Buehler_JAP}. Figure 2a)  shows the simultaneous response of both
rf-SETs in the time-domain to a 100kHz square wave signal of
amplitude $\mathrm{\sim 0.1e}$ applied to gate A2. The
capacitances between the gates and the SET islands were in the
range of $\mathrm{1-10aF}$ and the device resistances were
43.4k$\mathrm{\Omega}$ (SET1) and 30.4k$\mathrm{\Omega}$ (SET2).
Independent control of the dc source-drain bias across each SET
permits both devices to be simultaneously operated in the
superconducting state at the double Josephson quasi-particle
resonance, where near quantum-limited readout efficiency is approached \cite{Fulton_89,clerk_PRL}.
Figure 2b) shows the frequency spectrum of both
rf-SETs in response to an amplitude modulation (AM) gate signal
($\sim 0.1e$) at 2.5MHz. The frequency-domain sensitivity
($\mathrm{\delta q}$) of both devices was characterized by
measuring the signal to noise ratio of an AM side-band in a given
resolution bandwidth ($B$) \cite{aassime_APL}. For the case of
simultaneous operation, the minimum detectable signal was 
$\mathrm{\delta q = 7.5 \mu e/\sqrt{Hz}}$ for 
SET1 and $\mathrm{\delta q = 4.4 \mu e
/\sqrt{Hz}}$ for SET2. We note that determining the minimum
detectable charge $\mathrm{\Delta q = \delta q \times \sqrt{B}}$
from these sensitivity measurements assumes that the noise is white throughout the
measurement time. While this is true at higher frequencies (where
thermal noise from the rf amplifier and shot noise associated with
the SET current dominate), longer measurement times will include
unequally weighted or ``colored'' noise sources, typical of a $1/f$ spectrum.

In our setup the $1/f$ noise spectrum meets the white noise floor at 100kHz as shown in the inset to Fig 2 b). Consequently, 
measurements made on time-scales shorter than $10\mu s$ are not affected by $1/f$ noise. 
However, in the case of very small signals ($\mathrm{\Delta q <
0.01e}$) the measurement time required for high fidelity readout
increases drastically due to $1/f$ noise since the noise increases with 
$t_{meas}$ but signal only with $\sqrt{t_{meas}}$. This
limitation defines a lower limit (of $\mathrm{\Delta q \sim 0.01e}$) on the size of the charge signal
that can be read out efficiently in a single shot measurement. 

\begin{figure}[t]
\begin{center}
\includegraphics[width=8cm]{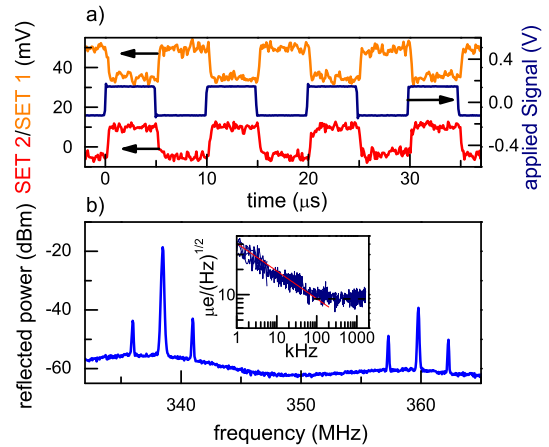}\vspace{-0.3cm}
\caption{\small{ \textbf{a)} Time-domain response of both rf-SETs to a   
100kHz square wave signal applied to gate A2 ($ \mathrm{\equiv   
0.1}$e)  \textbf{b)} The frequency-domain response of both rf-SETS to
a 2.5MHz AM signal (blue trace). We measure sensitivities of $\mathrm{\delta q = 7.5 \mu e/\sqrt{Hz}}$ for
SET1 and $\mathrm{\delta q = 4.4 \mu e
/\sqrt{Hz}}$ for SET2. \textbf{Inset:} Charge sensitivity as a function of frequency. Below 100kHz the sensitivity 
is limited by $1/f$ noise.}}\label{fig_td_sens} \vspace{-0.9cm}
\end{center}
\end{figure}

In order to explore the interplay between charge noise and readout
fidelity experimentally, we have performed single-shot time-domain
measurements of various signal amplitudes. The data in Figure 3 was obtained by applying a square wave signal to
gate G2 (see Figure 1). The trace in
Figure \ref{fig_td_sens}a) shows the result of 256 averages of a
$\mathrm{\Delta q = 0.01e}$ step. Figure 3b)
shows single-shot traces of a $\mathrm{\Delta q = 0.05e}$ and
$\mathrm{0.2e}$ gate signal in a measurement window of
$\mathrm{t_{meas}=0.6 \mu s}$. The slope of the transition between
the two levels is finite since the bandwidth of the resonant
circuit ($\mathrm{\sim 100 ns}$) acts as a low pass filter.
Figures 3(c-e) show the corresponding signal histograms and fitted
Gaussian probability distribution functions (PDFs) for the time-domain data. 
The overlap of the PDFs corresponds to the probability of mistaking one readout signal level for the other.
\begin{figure}[t]
\begin{center}
\includegraphics[width=8cm]{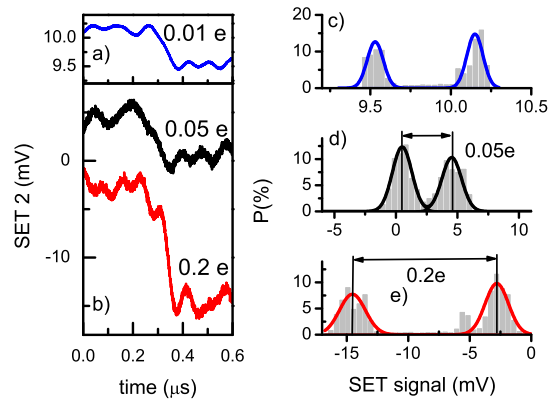}\vspace{-0.3cm}
\caption{\small{ \textbf{a)} Time-domain data for  $\mathrm{\Delta q = 0.01e}$,
obtained by averaging 256 times. \textbf{b)} Single-shot response of SET2
to a $\mathrm{\Delta q = 0.2e}$ and $\mathrm{0.05e}$ step in a
measurement time of $\mathrm{600ns}$.  \textbf{(c-e)} show the
corresponding probability distribution functions for each signal.
We measure a time-domain sensitivity of $\mathrm{5.3 \times
10^{-5}e/\sqrt{Hz}}$ in a bandwidth of 140 kHz for SET2.
}}\label{fig_td_sens} \vspace{-0.9cm}
\end{center}
\end{figure}

Although $1/f$ noise is negligible in the time considered here, we find additional noise sources that reduce the  measured time-domain
sensitivity (which is equal to the full width at half maximum of
the PDF for a SNR of 1), in comparison to measurements made in the
frequency-domain, where signals are strongly averaged (as in Fig. 2b) \cite{Dicke_1946}.
This difference in sensitivity can be explained by taking into account the additional noise introduced by 
the oscilloscope, demodulation circuitry and the effect of the
finite rf-SET bandwidth ($\mathrm{\sim }$10MHz) in comparison to
the measurement time  $\mathrm{t_{meas}} = 0.6 \mu s$. Our single-shot time-domain sensitivity ($\delta q = 5.3 \times 
10^{-5}e/\sqrt{Hz}$)
differs by close to a factor of 10 in comparison to our frequency-domain measurements ($\delta q = 4.4 \times 10^{-6}e/\sqrt{Hz}$ which are averaged) and 
highlights the need for a low-noise measurement set-up in addition to a near quantum-limited detector. In this regard we note
that the time-domain sensitivity is a more appropriate parameter characterizing the time required to perform a single-shot measurement.  

\begin{figure}
\begin{center}
\includegraphics[width=8cm]{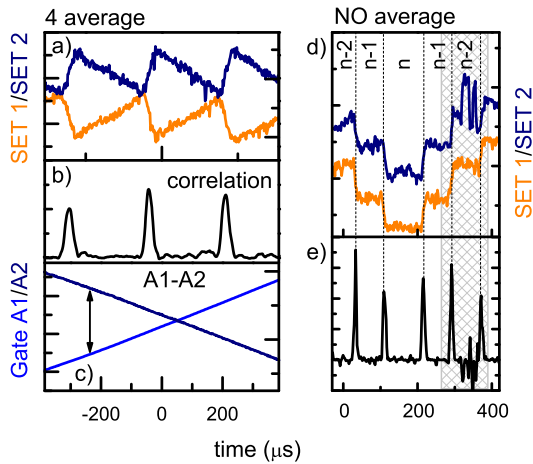}\vspace{-0.3cm}
\caption{\small{\textbf{a)}The simultaneous response of both SETs to
electron transfer between the two metal dots. \textbf{b)} Sharp peaks are produced
in the correlation whenever a transfer event occurs. \textbf{c)} shows the bias applied to gates A1 and A2 to create a differential
electric field. \textbf{d)} Single-shot time-domain measurements of charge
transfer in the double dot system. \textbf{e)} Correlation of the signals from
both SETs allows for error rejection in the event of charge noise
(shaded region). }}\label{fig_e_trans}\vspace{-0.9cm}
\end{center}
\end{figure}

In the above treatment we have neglected the effects associated
with telegraph noise originating from charge traps that are
strongly coupled to the SET. These large discontinuous {\it jumps}
have the potential to compromise readout in two ways: (1) high
frequency switching events that occur {\it during} the measurement
time greatly increase the time required to achieve high readout
fidelity and (2) slower fluctuations can produce readout errors if
they occur in the time interval between calibration and readout.
In this context calibration is the process of mapping the basis
states of the qubit to characteristic output signal levels of the
SET. This calibration process is essential and places additional
constraints on the requirement for detector stability.

In an effort to study the effect of telegraph noise on readout we
have used the twin rf-SET to measure the charge state of a double
dot system. In this experiment, electrons are adiabatically
transferred between two aluminum metal dots using an electric
field established by the adjacent electrodes (A1 and A2) shown in
Figure 1. This charge motion is simultaneously
detected by {\it two} rf-SETs located either side of the double
dot. Gates G1 and G2 (see Fig.1) are used to keep the rf-SETs biased to charge sensitive regions by nulling the direct effect of the A-gates 
on the SETs. For the architecture used here
the induced charge signals on the SET islands due to  charge
transfer in the double dot is of order $\mathrm{0.1
e}$~\cite{buehler_APL}. Cross-correlation of signals from the two
rf-SET permits suppression of spurious charge noise that
originates from fluctuating charge traps in the surrounding
material system. Although we have previously applied correlated
measurements to suppress the effect of telegraph noise in dc-SETs
\cite{buehler_APL}, the present work demonstrates our ability to
detect the motion of single electrons on microsecond time-scales
and distinguish, in real-time, transfer signals from background
charge noise.

Figure 4a) shows the simultaneous response of
both rf-SETs to electron motion between the two dots. The
characteristic sawtooth behavior arises from the polarization of
the metal dots by the electric field (from gates A1 and A2, see Fig.4c)), followed by a tunnel event
when the field overcomes the charging energy of the double dot
system. The signals are in anti-phase as one SET senses the
departure of an electron and the other its arrival. Using a
multichannel digital oscilloscope it is possible to multiply the
derivatives of the signals and thereby perform a real-time
cross-correlation. Sharp peaks are produced in the correlation
whenever a tunnel event occurs, as shown in Figure 4b).

Figure 4d) shows the result of a single-shot
measurement of the charge state of the double dot system. In this
instance telegraph noise affected the output of SET2 but not SET1
as indicated by the shaded region of the Figure. By comparing the
signals from both SETs we are able to register that an effective
readout error has occurred. We believe that such detection schemes
maybe of importance in maintaining high readout
fidelity, in the presence of charge noise. 

Charge noise is unavoidable in the solid state.
In systems where the rf-SET is used to perform readout, charge noise serves to 
reduce readout fidelity and limits the minimum charge signal detectable in a single-shot
measurement. We note that these constraints are an important consideration for the feasibility and design 
of readout architectures for both quantum computers and classical nano-scale devices. For architectures 
that have capacitive coupling to the readout detector similar to our nano-scale device, (such that $\Delta q >> 0.01$ see Fig.1b)) 
the influence of $1/f$ noise is negligible, provided readout can be performed on time-scales $<10\mu s$. Note that even for this case 
however, the presence of telegraph noise presents a challenge to high-fidelity readout. The 
cross-correlated single-shot measurements of a nano-scale device presented here,  serve to illustrate 
that this challenge can be overcome by correlating signals from two rf-SETs. 
In conclusion we have investigated the rf-SET as an efficient and
sensitive readout device capable of detecting the motion of single electrons on 
sub-microsecond time-scales, even in the presence of both $1/f$ and telegraph charge noise.

We thank D. Barber for technical support. This work was supported
by the  Australian Research Council, the Australian Government and
by the US National Security Agency (NSA), Advanced Research and
Development Activity (ARDA) and the Army Research Office (ARO)
under contract number DAAD19-01-1-0653. DJR acknowledges a
Hewlett-Packard Fellowship.

\end{document}